\documentclass[12pt]{article}
\usepackage{amssymb}
\usepackage{graphicx,color}

\begin{document}
\author{\large Seyen Kouwn$^1$\footnote{seyen@skku.edu}\,,
Joohan Lee$^2$\footnote{joohan@kerr.uos.ac.kr}\,,
Tae Hoon Lee$^3$\footnote{thlee@ssu.ac.kr}\,,
Taeyoon Moon$^4$\footnote{dpproject@skku.edu}\,
and Phillial Oh$^1$\footnote{ploh@skku.edu},
\\[3mm]
{\it  \small $^1$Department of Physics and
Institute of Basic Science, }\\
{\it \small Sungkyunkwan University, Suwon 440-746 Korea}
\\[3mm]
{\it  \small $^2$Department of Physics, University of Seoul, Seoul 130-743, Korea}
\\[3mm]
{\it  \small $^3$Department of Physics and Institute of Natural Sciences, }\\
{\it \small Soongsil University, Seoul 156-743 Korea}
\\[3mm]
{\it \small $^4$Center for Quantum Space-time, Sogang University, Seoul, 121-742, Korea}}

\title{Asymptotically Static  Universe}
\maketitle

\begin{abstract}
We consider a cosmology in which the final stage of the Universe is neither accelerating nor decelerating,
 but approaches an asymptotic state where the scale factor becomes a constant value.
In order to achieve this, we first bring in a scale factor with the desired property and then determine the details of the energy contents as a result of the cosmological evolution equations.
We show that such a scenario can be realized if we introduce a generalized quintom model which consists of a scalar field and a phantom with a {\it negative} cosmological constant term.
The standard cold dark matter with $w_m=0$ is also introduced. This is possible basically due to the balance between
the matter and the {\it negative} cosmological constant which tend
 to attract and scalar field and phantom which repel
in the asymptotic region. The stability analysis shows that this asymptotic solution is classically stable.
\end{abstract}
\newpage
\section{Introduction}

%Why it is necessary to introduce a phantom

 Recent observational data \cite{ob,Komatsu:2008hk} suggests that
 dark energy is believed to be responsible for the observed  acceleration of the Universe
 \cite{carroll}.
 Many theoretical proposals  on the nature of dark energy have been put forward.
 Cosmological constant \cite{weinberg,carol,sahni,peeb,padm} is widely accepted as the most promising candidate for dark energy, but there is the problem of
its extreme fine-tuning.
%Some dynamical model of cosmological constant, notably the quintessence model
%also requires a degree of fine tuning.
There are many alternative approaches which do not contradict with observational data \cite{amen}.
 One possible approach is based on the fact that the matter with the equation of state parameter $w < -1$
is not ruled out by observational data and moreover
necessary to
describe the current acceleration of the universe.
Phantom \cite{Caldwell:1999ew} is an alternative form of dark energy with this
property and
various aspects have been discussed extensively \cite{Caldwell:2003vq,Scherrer:2004eq,Kujat:2006vj,Barrow:2009sj}.

In spite of many theoretical attempts, at present
the nature of the dark energy is still enigmatic. An alternative approach to dark energy could be to try the
"reverse" way; Instead of guessing the energy density and then trying
to determine the evolution of the universe, one first assumes some
specific scale actor $a(t)$, and then fix the energy density from
the the equations of motion \cite{ellis,Capozziello:2005tf,Barrow:2004xh,Barrow:2004he}. This type of approach was proposed
before in which the focus has been primarily on the description of
the present universe starting from the past. In this work, we will
be interested in the future of the universe. So far, many results of the existing
literatures on the future of the Universe is based on the
consequences of acceleration \cite{Sami:2003xv,elizalde,barrow}.

However, the ignorance of the exact content of the dark energy leaves
other choices of the scale factor as viable. In this paper, we will
introduce a particular form of the scale factor which is consistent
with the present accelerating universe. It represents the universe which
does not expand forever, but approaches asymptotically a constant
value\footnote{The difference with the Einstein static universe is that in this case, the universe
stays static or starts from a static state \cite{ellis2}, whereas in our case the static state
is approached asymptotically in the future.
It can be shown that the Einstein static universe is stable against inhomogeneous perturbations
\cite{Harrison:1967zz,Gibbons:1987jt,Barrow:2003ni}, even though it
is unstable when the  homogeneous perturbations are performed.
We only consider homogeneous perturbations in this work.}.
The motivation for considering such Universe is that there is a priori no reason
why the acceleration of the universe should be the final state.
The $\Lambda-$CDM suggests the acceleration as a final stage,
but other contents of the dark energy could give different fate of the Universe.
For example, in the phantom model the energy density may increase with time and the Universe
ends up in big-rip singularity \cite{Caldwell:2003vq,Scherrer:2004eq,Chimento:2008ws} in which the
Hubble parameter and the curvature diverges in a finite time.
Another speculation is that dark energy might dissipate with time and
Universe contracts in a big-crunch~\cite{Khoury:2001bz,wang} as in the cyclic model~\cite{Steinhardt:2001vw,stein,Steinhardt:2004gk}.
Here, we investigate Universe neither accelerating nor decelerating,
but becoming asymptotically static with its size approaching a constant value, and
ask what type of energy contents will yield such Universe.
%We show that such a
%scenario is possible in the quintom model
%which consists of  a scalar filed and phantom.

In order to implement the idea,  we consider a generalized quintom model~\cite{Guo:2004fq,Zhao:2005vj,Cai:2009zp}
which consists of  a scalar field and phantom, and also the cold dark matter which satisfies the continuity equation. The potentials of the scalar fields and other quantities defining the model
 are not fixed at this  stage. With this, we bring in a scale factor which has the property that it is accelerating at present, but begins to decelerate after a fixed time, and eventually settles down to an asymptotic value. Then, the details of the theory are determined through the evolution equations.
We show that such a scenario is possible due to the balance between the attractive matter
and the negative cosmological constant, and the repulsive scalar field and phantom in the asymptotic region. The stability analysis shows that this asymptotic solution is classically stable.

%%%%%%%%%%%%%%%%%%%%%%%%%%%%%%%%%%%%%%%%%%%%%%%%%%%%%%%%%%%%%%%%%%%%%%%%%%%%%%%%%%%%%%%%%%%%%%%%%%%%%%%%%%%%%%%%%%%%%%%%%%%%%
%%% Section
%%%%%%%%%%%%%%%%%%%%%%%%%%%%%%%%%%%%%%%%%%%%%%%%%%%%%%%%%%%%%%%%%%%%%%%%%%%%%%%%%%%%%%%%%%%%%%%%%%%%%%%%%%%%%%%%%%%%%%%%%%%%%
\section{Phantom model}

Let us first  consider phantom
cosmology where gravity is coupled to a scalar field with a scalar
coupling function and a potential. This model was considered before
\cite{Capozziello:2005tf} where the possibility of phantom-non-phantom
transition occurs in early time as well as at late time.
In this paper, we go one step further and discuss the future of the
Universe with such a scalar coupling function.

The action is given by
\begin{eqnarray}
S=\int d^4 x \sqrt{-g} \Big\{ \frac{1}{2\kappa^2}R -\frac{1}{2}\tilde{\omega}(\varphi)\nabla_\mu\varphi \nabla^\mu\varphi -V(\varphi)  \Big\} + S_m \,,
\end{eqnarray}
where $\kappa^2 \equiv 8 \pi G_N$ is the gravitational coupling, $\tilde{\omega}(\varphi)$ can change continuously, and $V(\varphi)$ is the potential. $S_m$
is the matter action which we assume to be a cold dark matter with $w_m=0$,
and satisfies the continuity equation. The scalar field $\varphi$ is a function of time alone.
The evolution equations are given by
\begin{eqnarray}
\frac{3}{\kappa^2}H^2 &=& \rho_{\varphi}+\rho_m \,,\nonumber \\
-\frac{2}{\kappa^2}\dot{H} &=& p_{\varphi} + \rho_{\varphi} + \rho_m \,,\label{oneEinEq}
\end{eqnarray}
where $H=\dot a/a$ is the Hubble parameter, the energy density $\rho_{\varphi}$ and the pressure $p_{\varphi}$ are
\begin{eqnarray}
\rho_{\varphi} &=& \frac{1}{2}\tilde{\omega}(\varphi)\,\dot{\varphi}^2+V \,, \nonumber \\
p_{\varphi} &=& \frac{1}{2}\tilde{\omega}(\varphi)\,\dot{\varphi}^2-V \,,
\end{eqnarray}
and  the scalar field satisfies
\begin{eqnarray}
\tilde{\omega}(\varphi)\ddot{\varphi} + \frac{1}{2}\tilde{\omega}'(\varphi)\dot{\varphi}^2 + 3H\tilde{\omega}(\varphi)\dot{\varphi} + \frac{\partial V(\varphi)}{\partial \varphi} = 0 \,.
\end{eqnarray}

There could be many expressions of scale factor which has the desired property of transition from acceleration to deceleration.
In this work, we choose a special ansatz of the scale factor as follows;
\begin{eqnarray}\label{sola}
a(t) = a_{\infty}(1-Ae^{-B(t/t_0)^n}) \,,
\end{eqnarray}
where $A,B$  are constants, $t_0$ is the current age of the Universe,
$a_\infty$ is the asymptotic value, and  $n$ is arbitrary. This ansatz shows the behavior that the Universe accelerates
until some time (given below) and then deceleration takes place until it reaches an asymptotic value.
Therefore, it does not fit the whole history of the Universe, and we assume it is an effective scale factor
which describes the Universe only after the late-time acceleration has begun, $t_{tr} \simeq 0.5t_0$.
The transition from acceleration to deceleration will happen at the
time $t_*$ satisfying the following conditions:
\begin{eqnarray}
\ddot{a}(t)\Big|_{t=t_*} =0 \,, ~t_*/t_0= \sqrt[n]{\frac{n-1}{n B}}.
\end{eqnarray}
The condition that the deceleration must take place at later time restricts $n>1/(1-B)$.

With this scale factor, the Hubble parameter is given by
\begin{eqnarray}
H(t) = \frac{1}{t_0} \frac{ABn(t/t_0)^{n-1} e^{-B(t/t_0)^{n}} }{ 1-Ae^{-B(t/t_0)^n} } \,, \nonumber
\end{eqnarray}
and the matter density is given as
%\begin{eqnarray}
%\dot{H}(t) = \frac{1}{{t_0}^2} \frac{ABn(t/t_0)^{n-2} e^{-B(t/t_0)^{n}} }{ (1-Ae^{-B(t/t_0)^n})^2 }
%\biggr[ (n-1) - nB(t/t_0)^n - A(n-1)e^{-B(t/t_0)^n} \biggr] \,, \nonumber
%\end{eqnarray}
\begin{eqnarray}
\rho_m (t) =  \frac{\rho_0 (1-Ae^{-B})^3}{ (1-Ae^{-B(t/t_0)^n})^3 }  \,. \nonumber
\end{eqnarray}
With the ansatz as in  \cite{Capozziello:2005tf}
\begin{eqnarray}\label{solPhi}
\varphi=t \,,
\end{eqnarray}
the Einstein equations~(\ref{oneEinEq}) can be rewritten as
\begin{eqnarray}
\tilde{\omega} (t) &=&  -\frac{2}{\kappa^2}\dot{H}(t) - \rho_m(t)  \,, \nonumber \\
V(t) &=&  \frac{1}{\kappa^2} \left( 3H^2(t) + \dot{H}(t) \right) - \frac{1}{2}\rho_m(t)  \,. \nonumber
\end{eqnarray}

In order to determine the constant $A$ and $B$, we demand
\begin{eqnarray}
a(t_0) &=& a_{\infty}(1-Ae^{-B}) \equiv a_0\,, \nonumber \\
H(t_0) &=& \frac{1}{t_0} \frac{ABne^{-B}}{1-Ae^{-B}} \simeq \frac{1}{t_0}. \nonumber
\end{eqnarray}
Then, we get the following two relations
\begin{eqnarray}
A = ( 1-\frac{1}{m} )e^{B}\,,~~~B = \frac{1}{n(m-1)}   \nonumber
\end{eqnarray}
where $ m \equiv \frac{a_{\infty}}{a_0}$.

The free parameters are given by $m$ and $n$; $m$ determined the final size of the universe, and $n$
is related with the transition time from acceleration to deceleration.
They can be restricted by the observational data.
The effective equation of state parameter $w_\varphi$ is given by
\begin{eqnarray}\label{oneEff}
w_\varphi = \frac{p_\varphi}{\rho_\varphi} = \frac{\frac{\tilde{\omega}(t)}{2}-V(t)}{\frac{\tilde{\omega}(t)}{2}+V(t)}\,,
\end{eqnarray}
and at the current time we confront the equation of state parameter $w_\varphi$ with experimental uncertainty
($-1.14 \leq w_\varphi \leq -0.88$)~\cite{Komatsu:2008hk} which constrains the parameter $n$ and $m$.
The numerical analysis is given in Fig. 1 which shows that there exists allowed region within the experimental uncertainty.
Among the possible values, we choose a point and present numerical result in Fig. 2.
The result shows that at current time, the potential is dominant so that the Universe is accelerating.
Also the scalar function $\tilde{\omega}$ is slightly negative and it is in the phantom phase.
As it evolves, the deceleration begins to take over as the scalar function
$\tilde{\omega}(\varphi)$ (kinetic energy) reaches its near peak value.
Then, both the potential and the scalar function decrease rapidly and becomes negative.
However, the matter density stays always positive, decreasing as $1/a^3$, and asymptotically it
cancels exactly with the scalar contributions of the potential and the kinetic energy.
This exact cancelation renders the Universe asymptotically static.

In Fig.~\ref{fig02fig03}, we plot various quantities for values of $(m,n)=(3.5,2)$.

\begin{figure}[ht]
\begin{center}
\scalebox{1}[1]{\includegraphics{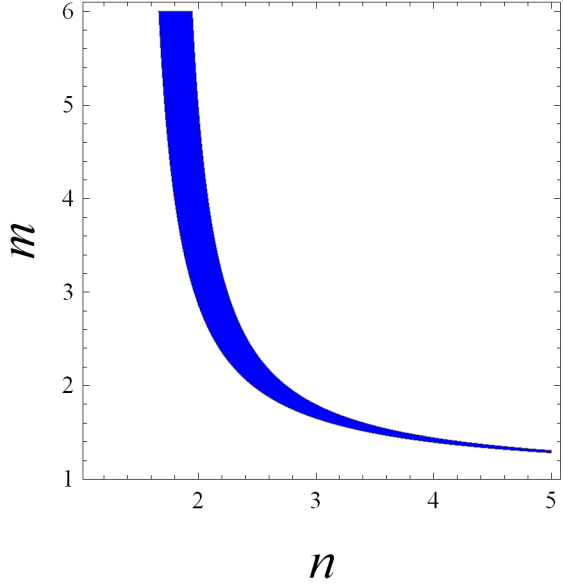}}
\end{center}
\caption{\small The experimental data constraints to the $(n,m)$ parameter space. The blue shaded region are allowed by
experimental uncertainty $-1.14 \leq w_\varphi \leq -0.88$. (We set the scale factor $a_0=1$, current time $t_0=1$, and $\kappa=1$.)}
\label{fig01}
\end{figure}

\begin{figure}[ht]
\begin{center}
\scalebox{1}[1]{\includegraphics{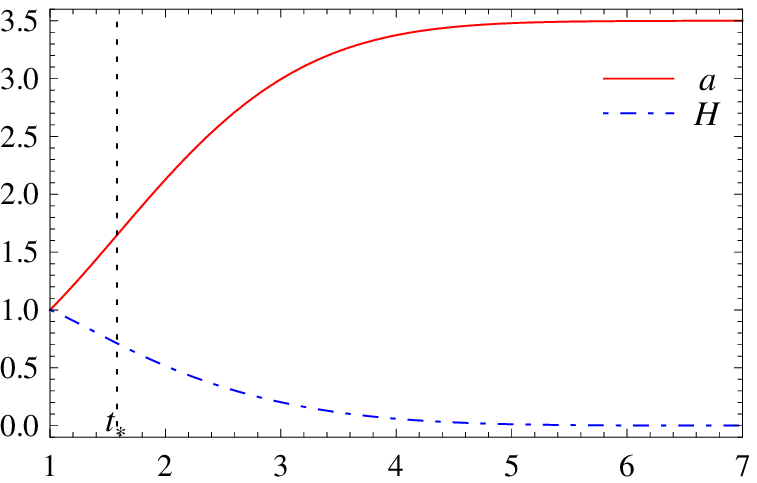} \includegraphics{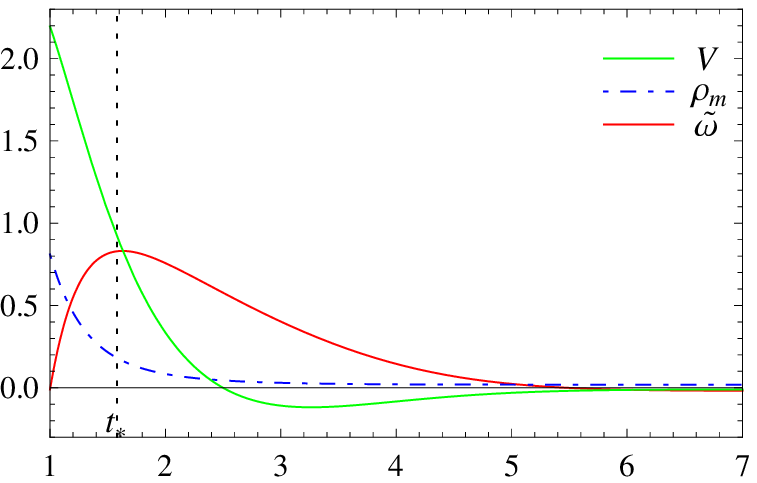}}
\end{center}
\caption{\small The left graphs are time evolution of the $a(t)~{\rm (solid~curve)}$ and $H(t)~{\rm (dash-dotted~curve)}$.
The right graphs are time evolution of the $V(t)~{\rm (green~solid~curve)}$, $\rho_m(t)~{\rm (dash-dotted~curve)}$,
and $\tilde{w}~{\rm (red~solid~curve)}$. (The parameter values are used for $n=2$ and $m=3.5$. We set the scale factor $a_0=1$,
current time $t_0=1$, and $\kappa=1$.)
In these figures, the Universe accelerates until the time  $t_* \equiv 1.58$.}
\label{fig02fig03}
\end{figure}

%%%%%%%%%%%%%%%%%%%%%%%%%%%%%%%%%%%%%%%%%%%%%%%%%%%%%%%%%%%%%%%%%%%%%%%%%%%%%%%%%%%%%%%%%%%%%%%%%%%%%%%%%%%%%%%%%%%%%%%%%%%%%
%%% Section
%%%%%%%%%%%%%%%%%%%%%%%%%%%%%%%%%%%%%%%%%%%%%%%%%%%%%%%%%%%%%%%%%%%%%%%%%%%%%%%%%%%%%%%%%%%%%%%%%%%%%%%%%%%%%%%%%%%%%%%%%%%%%
\section{STABILITY OF THE PHANTOM FIELD}

Let us check the stability of the ansatz  (\ref{sola}) and (\ref{solPhi}). In fact, this model( with one scalar field)
could be unstable due to several reasons. The first one is that the
effective equation of state parameter diverges, which occur in case of vanishing of denominator in (\ref{oneEff}).
The other one is the instability of the solution (\ref{sola}) and (\ref{solPhi})  due to singularity
which occurs when crossing the point of the coefficient of kinetic term $\tilde{\omega}(\varphi)$
becoming zero  as discussed in~\cite{Capozziello:2005tf}.
In order to avoid these problems, we may consider two scalar fields model like
\begin{eqnarray}
S=\int d^4 x \sqrt{-g} \Big\{ \frac{1}{2\kappa^2}(R-2\Lambda) -\frac{1}{2}\omega(\phi)\nabla_\mu\phi \nabla^\mu\phi -\frac{1}{2}\eta(\chi)\nabla_\mu\chi \nabla^\mu\chi -V(\phi,\chi)  \Big\} + S_m \,,
\end{eqnarray}
where $V(\phi,\chi)$ are potential energy for which we assume that there is no direct coupling between $\phi$ field and $\chi$ field, $V(\phi,\chi)=V_\phi(\phi) + V_\chi(\chi)$.
$\Lambda$ is a cosmological constant. For our purposes, we take the value of $\Lambda$ to be negative.

By varying the action, one can obtain the Einstein
equations and the continuity equation for CDM  as follows;
\begin{eqnarray}
\frac{3}{\kappa^2} H^2 &=& \rho_{\phi} + \rho_{\chi} + \rho_m + \frac{\Lambda}{\kappa^2} \,, \nonumber \\
-\frac{2}{\kappa^2} \dot{H} &=& \rho_{\phi} + \rho_{\chi} + p_{\phi} + p_{\chi} + \rho_m \,, \\
0 &=& \dot{\rho}_m + 3H \rho_m \,,
\end{eqnarray}
where energy density $\rho_{\phi}$, $\rho_{\chi}$ and pressure $p_\phi$, $p_\chi$ are given by
\begin{eqnarray}
\rho_{\phi} &=& \frac{1}{2}\omega(\phi)\,\dot{\phi}^2+V_\phi \,,~~~~\rho_{\chi} = \frac{1}{2}\eta(\chi)\,\dot{\chi}^2+V_\chi \,, \nonumber \\
p_{\phi} &=& \frac{1}{2}\omega(\phi)\,\dot{\phi}^2-V_\phi \,,~~~~p_{\chi} = \frac{1}{2}\eta(\chi)\,\dot{\chi}^2-V_\chi \,,
\end{eqnarray}
and equations of scalar fields $\phi$ and $\chi$ as
\begin{eqnarray}\label{twoScaEq}
0 &=& \omega(\phi)\ddot{\phi} + \frac{1}{2}\omega'(\phi)\dot{\phi}^2 + 3H\omega(\phi)\dot{\phi} + \frac{\partial V_\phi(\phi)}{\partial \phi} \,, \nonumber \\
0 &=& \eta(\chi)\ddot{\chi} + \frac{1}{2}\eta'(\chi)\dot{\chi}^2 + 3H\eta(\chi)\dot{\chi} + \frac{\partial V_\chi(\chi)}{\partial \chi} \,.
\end{eqnarray}

\begin{figure}[ht]
\begin{center}
\scalebox{1}[1]{\includegraphics{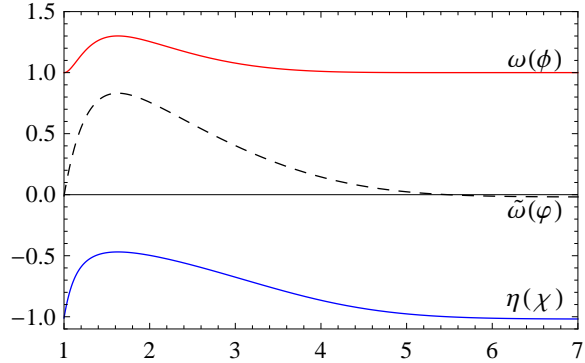}}
\end{center}
\caption{\small The graphs are evolution of the coefficient of the kinetic term $\omega(\phi)$ and $\eta(\chi)$.
The dashed curve is coefficient of the kinetic term $\tilde{\omega}(\varphi)$ of the one scalar field model.
(We set the parameter value $\alpha=1$).}
\label{fig04}
\end{figure}

In the two scalar fields model, we can choose the corresponding coefficients $\omega$ and $\eta$ to be non-vanishing everywhere.
Then  the instability due to divergence would not occur.
For example, we may choose that $\omega$ should be always positive and $\eta$ be always negative \cite{Capozziello:2005tf},
\begin{eqnarray}
\omega(\phi) &=&  \sqrt{\alpha^2 + \tilde{\omega}(\phi)^2} \,, \nonumber \\
\eta(\chi) &=&  \tilde{\omega}(\chi) - \sqrt{\alpha^2 + \tilde{\omega}(\chi)^2} \,,
\end{eqnarray}
where $\alpha$ is a  non-zero constant. Since $\omega$ and $\eta$ are positively- and negatively-defined, respectively,
$\omega$ gives ordinary scalar phase and $\eta$  phantom phase. The numerical evaluation is shown in Fig.~\ref{fig04}.
The corresponding potential energy $V_\phi(\phi)$ and $V_\chi(\chi)$
are obtained by solving the following differential equations that are just the scalar field equations~(\ref{twoScaEq}) with solution $\phi=t$ and $\chi=t$
\begin{eqnarray}
\dot{V}_\phi &=&  -\frac{1}{2}\dot{\omega} - 3H \omega \,, \nonumber \\
\dot{V}_\chi &=&  -\frac{1}{2}\dot{\eta} - 3H \eta \,,
\end{eqnarray}
where the solutions $V_\phi$ and $V_\chi$ are generally dependent upon integration constant which can be defined in accordance with
the potential energy of the one scalar fields model.

\begin{figure}[ht]
\begin{center}
\scalebox{1}[1]{ \includegraphics{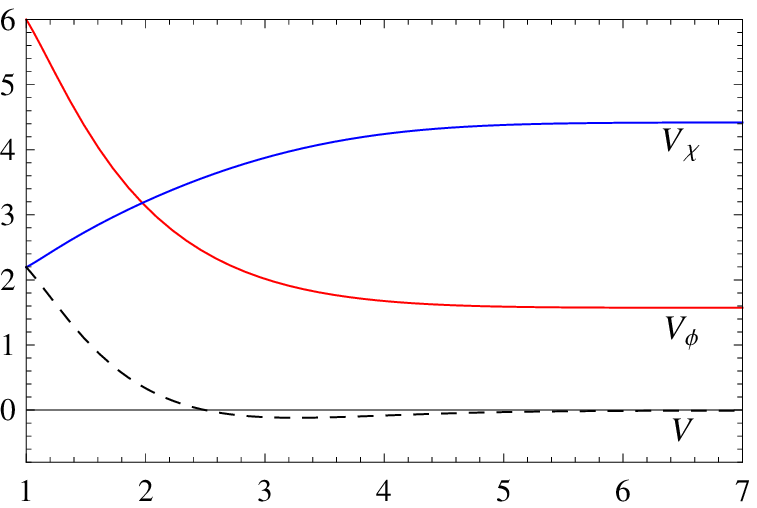}~~\includegraphics{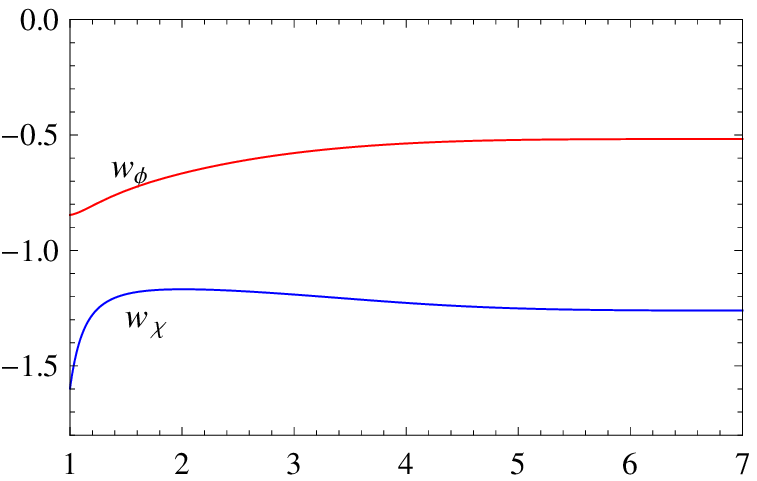}}
\end{center}
\caption{\small The left graphs are evolution of the potential energy of $V_\phi$~(red~curve) and $V_\chi$~(blue~curve).
The dashed curve is evolution of the potential energy of the one scalar field model. The right graphs are evolution of
the effective equation of state of field $\phi$~(red~curve) and field $\chi$~(blue~curve).
(We choosing the initial condition $V_\phi(1)=6$ and $V_\chi(1)=V(1)-V_\phi(1)-\Lambda/\kappa^2$ with $\Lambda=-6$ and $\kappa^2=1$.)}
\label{fig0506}
\end{figure}

The corresponding effective equation of state parameter $w_\phi$ and $w_\chi$ are given by
\begin{eqnarray}\label{twoEffPara}
w_\phi = \frac{p_\phi}{\rho_\phi} = \frac{\frac{\omega(t)}{2}-V_\phi(t)}{\frac{\omega(t)}{2}+V_\phi(t)}\,,~~~
w_\chi = \frac{p_\chi}{\rho_\chi} = \frac{\frac{\eta(t)}{2}-V_\chi(t)}{\frac{\eta(t)}{2}+V_\chi(t)}\,.
\end{eqnarray}
As was mentioned before, the effective equation of state parameter of the one scalar field model becomes singular at some point.
In order to obtain the well defined effective equation of state parameter of the two scalar field model,
we can choose the integration constant such that denominator in~(\ref{twoEffPara}) should be always non-vanishing.
This can be done and some numerical result is shown in Fig.~\ref{fig0506}
%%%%%

%%%%%

To gain  some insight into the property of the solution, we introducing the following variables
 \cite{Capozziello:2005tf}:
\begin{eqnarray}
X_\phi \equiv \dot{\phi}\,,~X_\chi \equiv \dot{\chi} \,,
\end{eqnarray}
and we consider the perturbations of variables $X_\phi\,,X_\chi\,,H\,,$ and $\rho_m$ near the background solution.
One can obtain the equations for the linear perturbations up to the first order as follows:
\begin{eqnarray}\label{varEqSol}
\frac{d}{dt} \delta X_{\phi} &=& \left( -\frac{\omega'}{\omega} -3H \right) \delta X_\phi -3 \delta H \,, \nonumber \\
\frac{d}{dt} \delta X_{\chi} &=& \left( -\frac{\eta'}{\eta} -3H \right) \delta X_\chi -3 \delta H \,, \nonumber \\
\frac{d}{dt} \delta H &=& -\omega \delta X_\phi -\eta \delta X_\chi -\frac{1}{2}\delta \rho_m \,, \nonumber \\
\frac{d}{dt} \delta \rho_m &=& -3 \rho_m \delta H -3H \delta \rho_m \,.
\label{variation}
\end{eqnarray}
Before we carry out the numerical analysis, let us analyze the late time property analytically. When time $t$ becomes very large,
the contribution of $H\,,\omega'\,,$ and $\eta'$ will become negligible. Therefore Eq.(\ref{variation}) reduces to
\begin{eqnarray}\label{tinfM}
\frac{d}{dt}
\left(
  \begin{array}{c}
    \delta X_{\phi}  \\
    \delta X_{\chi}  \\
    \delta H  \\
    \delta \rho_m  \\
  \end{array}
\right)
=M
\left(
  \begin{array}{c}
    \delta X_{\phi}  \\
    \delta X_{\chi}  \\
    \delta H  \\
    \delta \rho_m  \\
  \end{array}
\right)
\,,~~
M \equiv \left(
  \begin{array}{cccc}
    0 & 0 & -3 & 0 \\
    0 & 0 & -3 & 0 \\
    -\omega_\infty & -\eta_\infty & 0 & -\frac{1}{2} \\
    0 & 0 & -\rho_{m,\infty} & 0 \\
  \end{array}
\right) \,,
\end{eqnarray}
where $\omega_\infty\,,\eta_\infty$ and $\rho_{m,\infty}$ are values at late time. The eigenvalues of matrix $M$ are given by
\begin{eqnarray}\label{varEqSol}
0\,,~0\,,\pm\sqrt{3\left(\omega_{\infty}+\eta_{\infty}+\frac{1}{2}\rho_{m,\infty}\right)} \,.
\end{eqnarray}

\begin{figure}[ht]
\begin{center}
\scalebox{0.55}[0.55]{\includegraphics{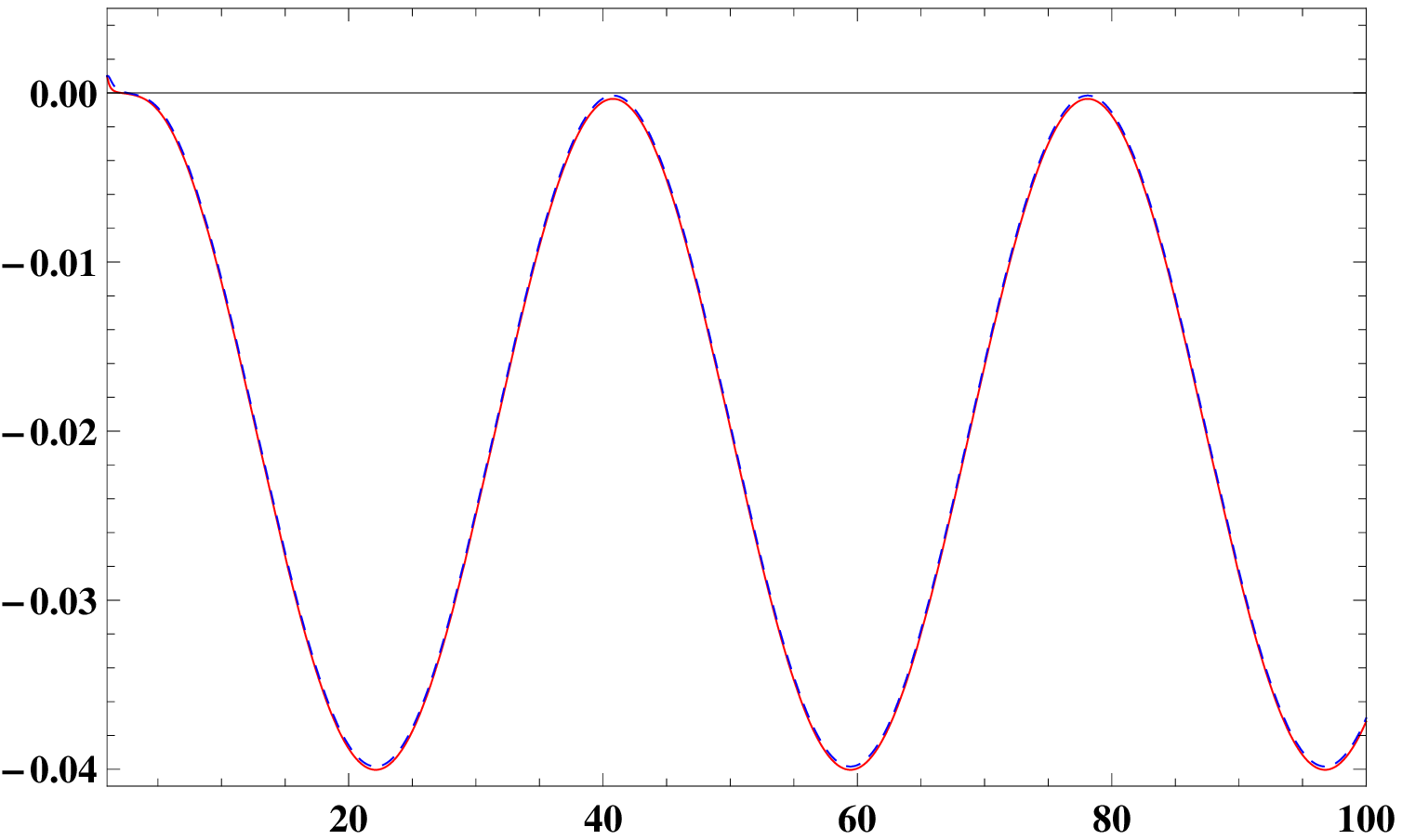} \includegraphics{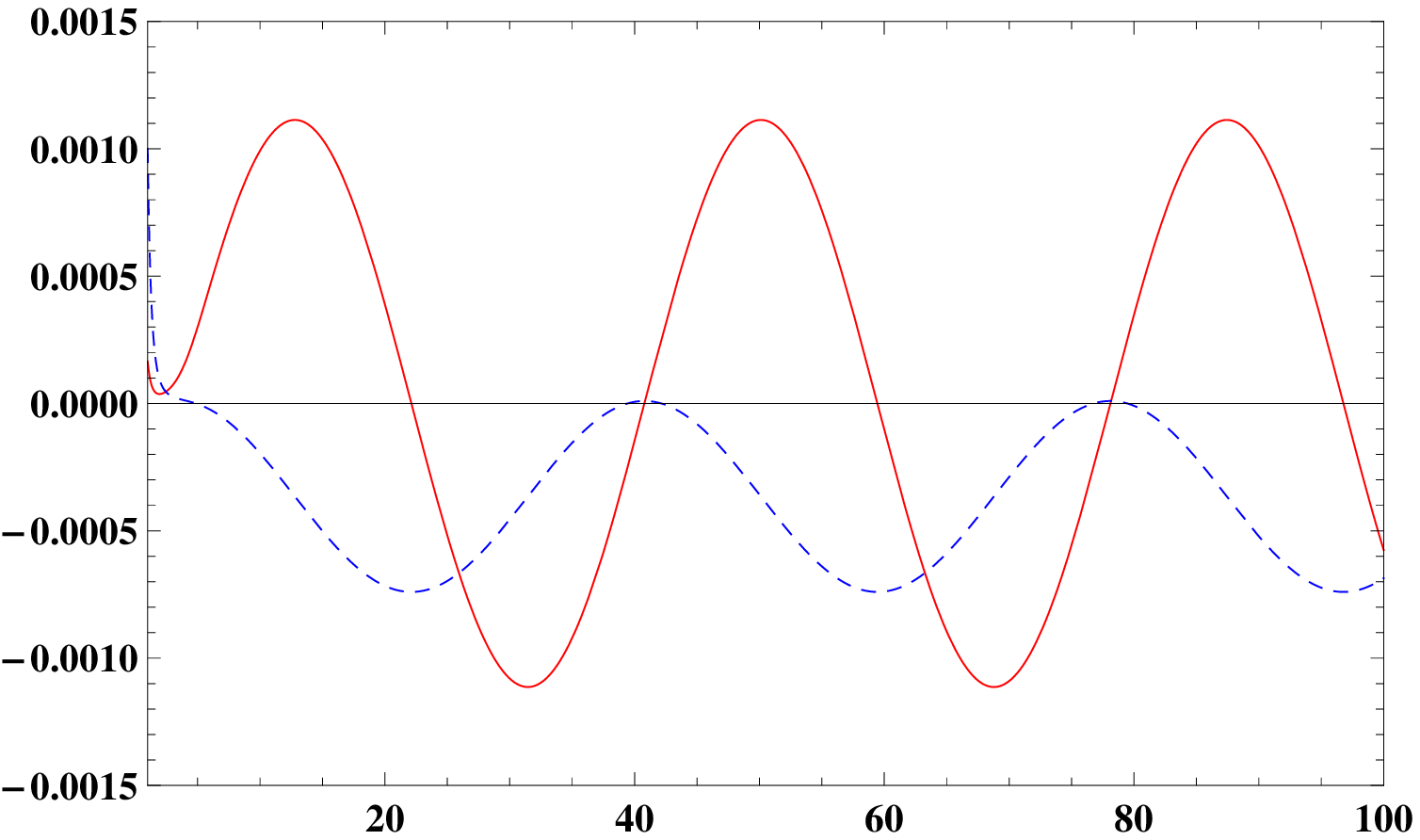}}
\end{center}
\caption{\small The left graphs are time evolution of the $\delta X_\phi~{\rm (solid~curve)}$ and $\delta X_\chi~{\rm (dashed~curve)}$.
The right graphs are time evolution of the $\delta H~{\rm (solid~curve)}$ and $\delta \rho_m~{\rm (dashed~curve)}$.
(The parameter values used are $n=2\,,m=3.5$ and $\alpha=1$ and computation done by choosing the initial conditions
$\delta X_{\phi}=0.001, \delta X_{\chi}=0.001, \delta H=0.000165$, and $\delta \rho_m =0.001$.)}
\label{varSol}
\end{figure}

Since $\omega_{\infty}+\eta_{\infty}+\frac{1}{2}\rho_{m,\infty} \simeq -\frac{\rho_0}{2m^3}$ is negative(for the positive value $m$),
there are no positive eigenvalues. Thus, the solution is stable in the late time.
We mention that this result can also be obtained by solving the differential equations Eq.(\ref{tinfM}), and in the late time region, the solutions are
\begin{eqnarray}\label{lateSol}
\delta X_{\phi}(t) &=& A_\phi + B \cos(\lambda t) + C \sin(\lambda t) \,, \nonumber \\
\delta X_{\chi}(t) &=& A_\chi + B \cos(\lambda t) + C \sin(\lambda t) \,, \nonumber \\
\delta H(t) &=& \frac{\lambda}{3} \biggr( -C \cos(\lambda t) + B \sin(\lambda t) \biggr) \,, \nonumber \\
\delta \rho_m(t) &=& A_{\rho_m} + \rho_\infty \left( B \cos(\lambda t) + C \sin(\lambda t) \right) \,,
\end{eqnarray}
where $\lambda$, $A_\phi$, $A_\chi$, $A_{\rho_m}$, $B$, and $C$ are given by
\begin{eqnarray}
\lambda &=& \sqrt{-3\left(\omega_\infty + \eta_\infty +\frac{1}{2}\rho_{m,\infty} \right)} \,, \nonumber \\
A_{\phi} &=& \frac{3}{2\lambda^2} \biggr( I_4 + 2(I_2-I_1)\eta_\infty - I_1 \rho_{m,\infty} \biggr) \,, \nonumber \\
A_{\chi} &=& \frac{3}{2\lambda^2} \biggr( I_4 - I_2 \rho_\infty + 2(I_1-I_2)\omega_\infty  \biggr) \,, \nonumber \\
A_{\rho_m} &=& \frac{3}{2\lambda^2} \biggr( 2(I_2 \eta_\infty + I_1 \omega_\infty)\rho_{m,\infty} - 2 I_4 ( \eta_\infty + \omega_\infty) \biggr) \,, \nonumber \\
B &=& -\frac{3}{2\lambda^2} \biggr( I_4 + 2 I_2 \eta_\infty + 2 I_1 \omega_\infty  \biggr) \,, \nonumber \\
C &=& -\frac{3 I_3}{\lambda} \,,
\end{eqnarray}
with $I_1$, $I_2$, $I_3$, and $I_4$ being integration constants. We find that the solutions has an oscillating behavior
with finite amplitude at late times. A direct numerical analysis of the Eq. (\ref{variation}) also confirms the stability.
The results are shown in Fig.~\ref{varSol}. It shows that the perturbation settles down to
the one given in Eq. (\ref{lateSol}) very rapidly. The subsequent variation  oscillates around the solution.

\section{Conclusion}

In this paper,  we considered a generalized quintom model
which consists of  a scalar field and phantom, and also the cold dark matter which satisfies the continuity equation.
The potential of the scalar fields and other quantities defining the model
 are determined by the condition that they are consistent with the scale factor which has the property
 that it is accelerating at present, but begins to decelerate after a fixed time, and eventually settles down to an asymptotic value.
 The details of the theory are determined through the evolution equations.
We showed that the asymptotic scenario is possible due to the balance between the matter
and the negative cosmological constant which tend
 to attract and scalar field and phantom which repel
in the asymptotic region.
The stability analysis shows that this asymptotic solution is classically stable.

The scale factor  showing the behavior of approaching an asymptotic state is by no means unique.
%%% from mail
We could try other ansatz which exhibits similar property
with the one investigated in this paper. Obviously, for different choices of the scale factor,
the energy contents will be given differently.
Nevertheless, the over-all feature of necessity of the phantom will remain,
because in order to achieve the asymptotic final stage,
the attractive force of the matter must balance the repulsive phantom.
It remains to be seen whether the exact nature of the dark energy could support
such an asymptotically static Universe.

%It is to be reminded that the realistic scale factor awaits be
%selected only after  the exact nature of the dark energy can be identified.
%Until then, many possibilities remain as long as they do not contradict with the observational data today.

\section{Acknowledgments}

This work was supported by the Basic Science Research Program through the National Research
Foundation of Korea (NRF) funded by the MEST (2011-0026655) and
by NRF grant funded by the
Korea government(MEST)through the Center for Quantum Spacetime (CQUeST) of
Sogang University with grant number 2005-0049409.

\end{document}